\documentclass[11pt,twocolumn,twoside]{article}
\usepackage{iccr2024}
\usepackage{amsfonts}
\usepackage{tabularx}
\usepackage{svg}
\usepackage{subcaption} 
\usepackage{float}

\begin{document}

\title{Diffusion Schrödinger Bridge Models for High-Quality MR-to-CT Synthesis for Head and Neck Proton Treatment Planning} 

\author[1,2]{Muheng Li}
\author[1,3]{Xia Li}
\author[1]{Sairos Safai}
\author[1,4,5]{Damien Weber}
\author[1,2]{Antony Lomax}
\author[1]{Ye Zhang}

\affil[1]{Center for Proton Therapy, Paul Scherrer Institut, Villigen}

\affil[2]{Department of Physics, ETH Zurich, Zurich}

\affil[3]{Department of Computer Science, ETH Zurich, Zurich}

\affil[4]{Department of Radiation Oncology, University Hospital of Zurich, Zurich}

\affil[5]{Department of Radiation Oncology, Inselspital, Bern University Hospital, University of Bern, Bern}

\maketitle
\thispagestyle{fancy}


\begin{customabstract}
    In recent advancements in proton therapy, MR-based treatment planning is gaining momentum to minimize additional radiation exposure compared to traditional CT-based methods. This transition highlights the critical need for accurate MR-to-CT image synthesis, which is essential for precise proton dose calculations. Our research introduces the Diffusion Schrödinger Bridge Models (DSBM), an innovative approach for high-quality MR-to-CT synthesis. DSBM learns the nonlinear diffusion processes between MR and CT data distributions. This method improves upon traditional diffusion models by initiating synthesis from the prior distribution rather than the Gaussian distribution, enhancing both generation quality and efficiency.  We validated the effectiveness of DSBM on a head and neck cancer dataset, demonstrating its superiority over traditional image synthesis methods through both image-level and dosimetric-level evaluations. The effectiveness of DSBM in MR-based proton treatment planning highlights its potential as a valuable tool in various clinical scenarios.
    \end{customabstract}


\section{Introduction}

In radiation therapy, proton therapy stands out for its exceptional precision and ability to minimize radiation exposure to non-target tissues. In traditional proton therapy treatment planning, reliance on Computed Tomography (CT) scans is standard practice. However, CT imaging for pre-treatment planning, as well as daily verification, often results in extra radiation doses to the patient. Recognizing this challenge, there's a growing interest in exploring Magnetic Resonance (MR) imaging as a potential alternative, given its ability to provide detailed anatomical information without the associated radiation exposure \cite{hoffmann2020mr}. This paradigm shift towards MR-based proton treatment planning underscores the need for high-quality MR-to-CT image synthesis, where attaining both anatomical precision and accurate density information within synthetic CTs is essential for proton dose calculations.

Recent advancements in AI-based generative imaging techniques, particularly the Diffusion Models (DMs) \cite{ho2020denoising}, have shown promising results for natural image generation tasks \cite{dhariwal2021diffusion}. DMs have also been proven for their capacity to produce detailed and realistic synthetic medical images \cite{khader2023denoising}. However, traditional DMs often involve lengthy sampling chains, which can significantly reduce sampling efficiency. Furthermore, given the higher image generation variability compared to conventional approaches, DMs may have limitations in clinical MR-to-CT synthesis, where both image generation precision and stability are crucial.

To tackle these challenges, we propose the Diffusion Schrödinger Bridge Model (DSBM) for MR-to-CT synthesis. Unlike conventional diffusion methods that sample from a Gaussian distribution, this approach directly learns the nonlinear diffusion processes between MR and CT data distributions. This method can establish a more direct transition model between MR and CT data, enhancing image generation quality and stability while increasing efficiency through fewer sampling steps, thereby boosting its potential application in proton therapy clinical practices.


\section{Materials and Methods}

\subsection{Diffusion Schrödinger Bridge Models (DSBM)}

Schrödinger Bridge (SB), as introduced by Schrödinger in 1932 \cite{schrodinger1932theorie}, represents an entropy-regularized optimal transport model that incorporates the following forward-backward Stochastic Differential Equations (SDEs):
\begin{subequations}
\begin{align}
& \mathrm{d} X_t=\big[f_t+g^2_t \nabla \log \Psi_t\big(X_t\big)\big] \mathrm{d} t+g_t \mathrm{~d} W_t, \label{eq:1a} \\
& \mathrm{d} X_t=\big[f_t-g^2_t \nabla \log \widehat{\Psi}_t\big(X_t\big)\big] \mathrm{d} t+g_t \mathrm{~d} \bar{W}_t, \label{eq:1b}
\end{align}
\end{subequations}
where $X_t \in \mathbb{R}^d$ is a $d$-dimensional stochastic process, indexed by discrete time steps $t\in[0,T]$. The Wiener process is represented as $W_t$, while its reverse-time process is denoted as $\bar{W}_t$. 
Here, $X_0$ is sampled from the boundary distribution $p_\mathrm{CT}$, representing CT image data, while $X_T$ is drawn from $p_\mathrm{MR}$, representing the paired MR image data. The functions $\Psi_t, \widehat{\Psi}_t$ represent time-dependent energy potentials that satisfy the following coupled Partial Differential Equations (PDEs):
\begin{equation}
\begin{gathered}
\left\{\begin{array}{l}
\frac{\partial \Psi_t(x)}{\partial t}=-\nabla \Psi^{\top} f-\frac{1}{2} g^2 \nabla^2 \Psi \\[3pt]
\frac{\partial \widehat{\Psi}_t(x)}{\partial t}=-\nabla \cdot(\widehat{\Psi} f)+\frac{1}{2} g^2 \nabla^2 \widehat{\Psi}
\end{array}\right. \\
\text { s.t. } \Psi_0(x) \widehat{\Psi}_0(x)=p_{\mathrm{CT}}(x), \Psi_T(x) \widehat{\Psi}_T(x)=p_{\mathrm{MR}}(x)
\end{gathered}  \label{eq:1}
\end{equation}
However, addressing the SB models involves simulating stochastic processes and requires resource-intensive iterative procedures \cite{de2021diffusion, chen2021likelihood}. Consequently, this approach faces challenges in terms of tractability and practical applicability.

To tackle this problem, we implement the concepts from \cite{liu20232}, transforming the SB model into the following SDEs:
\begin{subequations}
\begin{align}
\mathrm{d} X_t=f_t\left(X_t\right) \mathrm{d} t+g_t \mathrm{~d} W_t, X_0 \sim \widehat{\Psi}_0, \\
\mathrm{d} X_t=f_t\left(X_t\right) \mathrm{d} t+g_t \mathrm{~d} \bar{W}_t, X_T \sim \Psi_T.
\end{align}
\end{subequations}
This form adheres to the fundamental structure of the score-based generative models (SGM) \cite{song2020score}, where $\nabla \log \widehat{\Psi}_t\left(X_t\right)$ and $\nabla \log \Psi_t\left(X_t\right)$ are the \textit{score functions} of the above SDEs. 
Then, we delve into practical strategies for applying the above model to MR-to-CT synthesis. Given the fact that the availability of paired information during training, where $p\left(X_0, X_T\right)=p_{\mathrm{CT}}\left(X_0\right) p_{\mathrm{MR}}\left(X_T \mid X_0\right)$, with $p_{\mathrm{CT}}$ and $p_{\mathrm{MR}}$ corresponding to the CT and MR data distributions respectively. This framework allows us to construct tractable Schrödinger Bridges between $X_0$ and $p_{\mathrm{MR}}\left(X_T \mid X_0\right)$. According to 
\cite{liu20232}, we can adopt an analytic posterior given the boundary pair MR-CT data $(X_0,X_T)$:
\begin{equation}
\begin{gathered}
q\left(X_t \mid X_0, X_T\right)=\mathcal{N}\left(X_t ; \mu_t\left(X_0, X_T\right), \Sigma_t\right), \\
\mu_t=\frac{\bar{\sigma}_t^2}{\bar{\sigma}_t^2+\sigma_t^2} X_0+\frac{\sigma_t^2}{\bar{\sigma}_t^2+\sigma_t^2} X_T, \quad \Sigma_t=\frac{\sigma_t^2 \bar{\sigma}_t^2}{\bar{\sigma}_t^2+\sigma_t^2} \cdot I.
\end{gathered}
\end{equation}
Here, we define $\sigma_t^2=\int_0^t g^2_\tau \mathrm{d} \tau$, $\bar{\sigma}_t^2=\int_t^1 g^2_\tau \mathrm{d} \tau$. Then, based on the network parameterization settings in \cite{dhariwal2021diffusion} with a 3D U-Net \cite{cciccek20163d}, the training and sampling procedures can be done under the same framework as \cite{liu20232}.

\subsection{Dataset and Model Implementations}
This study used a dataset of head-and-neck images from 58 patients treated at PSI between 2017 and 2022, divided into training (46 patients) and testing (12 patients) sets. CT scans were performed using a Siemens Sensation Open CT scanner (resolution: 1 × 1 × 2 mm), and MR scans with a 1.5T Siemens Aera MR scanner (voxel size: 1 × 1 × 1 mm), primarily using the T1 vibe Dixon sequence. Data pre-processing, including a neural network-based data registration step, followed protocols outlined in \cite{li2024unified}. We strictly adhered to ethical standards in human data research, obtaining informed consent for using anonymized patient data and ensuring confidentiality and ethical compliance. DSBM was comprehensively compared with two other established medical image synthesis approaches, namely the pix2pix model \cite{isola2017image} and the nnU-Net model \cite{isensee2018nnu}. While the nnU-Net model utilized the same 3D data inputs as DSBM, the pix2pix model was trained on 2D image slices.

\subsection{Image-level Evaluation}

We evaluated the accuracy and anatomical fidelity of synthetic CT (sCT) and planning CT (pCT) using various image-related metrics, including: Mean Absolute Error (MAE) on Hounsfield Units (HU), Peak Signal-to-Noise Ratio (PSNR), and Structural Similarity Index (SSIM) for intensity value mapping, focusing on different regions of interest (RoIs). For anatomical conformity, Dice Coefficients (Dice) were calculated to assess structural overlaps, especially in bone regions (HU greater than 250).

\subsection{Dosimetric-level Evaluation}

Dosimetric evaluations were performed to validate the clinical utility of the generated sCT images in proton therapy. Treatment plans were optimized on pCT using in-house developed Juliana software \cite{bellotti2023juliana}, and doses were recalculated for sCTs as direct comparison. We assessed Mean Dose Error at various dose thresholds (10\%, 50\%, and 90\%), and used gamma pass rates (1\%/1mm and 2\%/1mm) for gamma analysis between planned and recalculated dose distributions. Dose-Volume Histograms (DVHs) and Differential Dose-Volume Histograms (DDVHs) were analyzed, providing insights into the dosimetric impact of using sCTs. We further investigated the variations in several typical dose indices, specifically V95, D95, D2, Dmean, and Vdosediff>3\% (the percentage of voxels in the same structure with an absolute point dose difference larger than 3\% of the prescribed dose), for both PTVs and OARs, to comprehensively understand the accuracy of the new MR-to-CT synthesis approach for calculating proton therapy treatment plans.

\section{Results}

\begin{figure*}
    \centering
    \begin{subfigure}[b]{0.5\linewidth}
        \includegraphics[width=\linewidth]{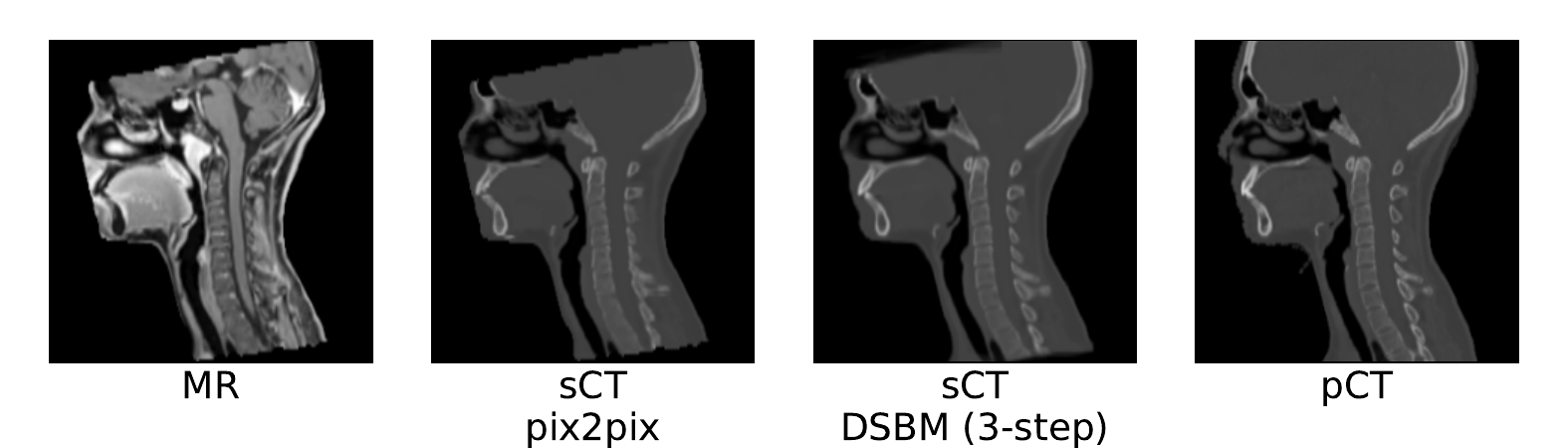}
    \end{subfigure}
    \begin{subfigure}[b]{0.49\linewidth}
        \includegraphics[width=\linewidth]{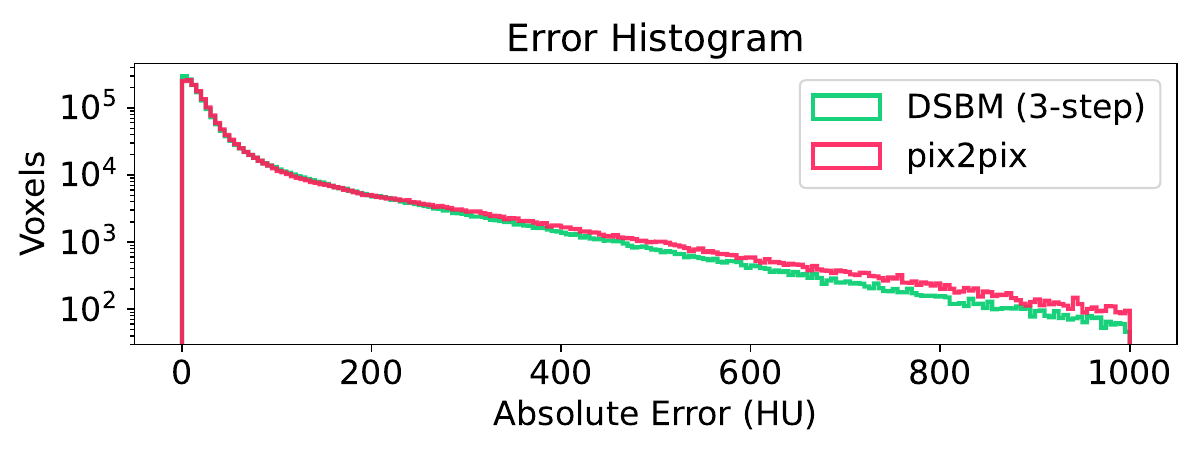}
    \end{subfigure}
    \caption{Visual comparisons of sCTs and error histograms from pix2pix and DSBM. Errors are calculated as $|sCT - pCT|$.}
    \vspace{-5pt}
    \label{fig:combined}
\end{figure*}

At the image level, DSBM outperformed conventional nnU-Net and pix2pix models in the testing set. It achieved a lower Mean Absolute Error (MAE) of 72.30 ± 8.31 HU and a higher Dice score of 83.10 ± 3.90\%, indicating enhanced anatomical accuracy, especially in bone structures (Table \ref{tab:1}). Visual comparisons in Figure \ref{fig:combined} further corroborate this, with error histograms showing DSBM's reduced errors and improved fidelity in synthetic CT images (sCTs).

\begin{figure*}
  \centering
  \begin{subfigure}{1.0\textwidth}
    \includegraphics[width=\linewidth]{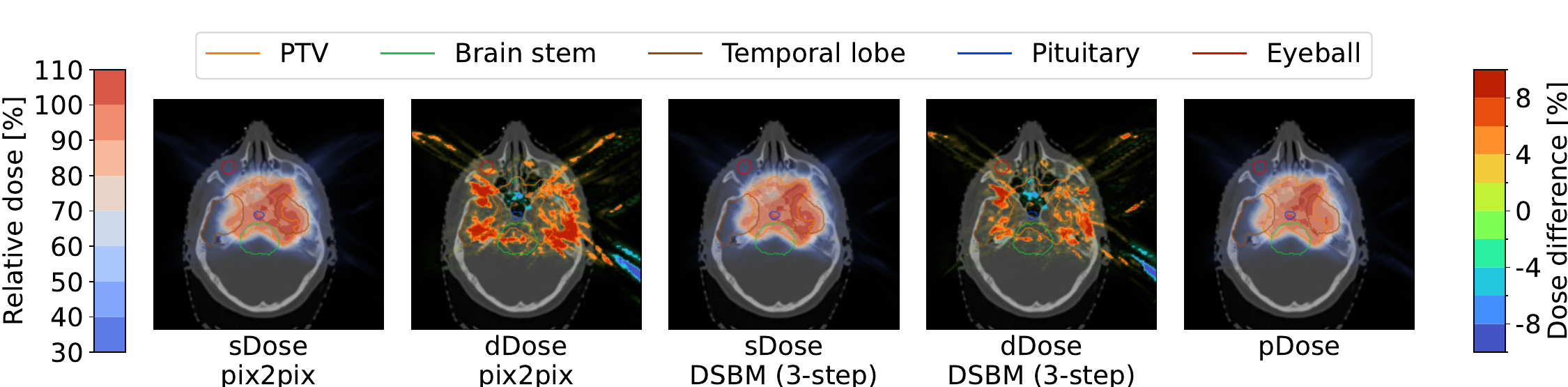}
    \label{fig:sub1}
  \end{subfigure}
  
  \vspace{-10pt} 
  \begin{subfigure}{0.5\textwidth}
    \includegraphics[width=\linewidth]{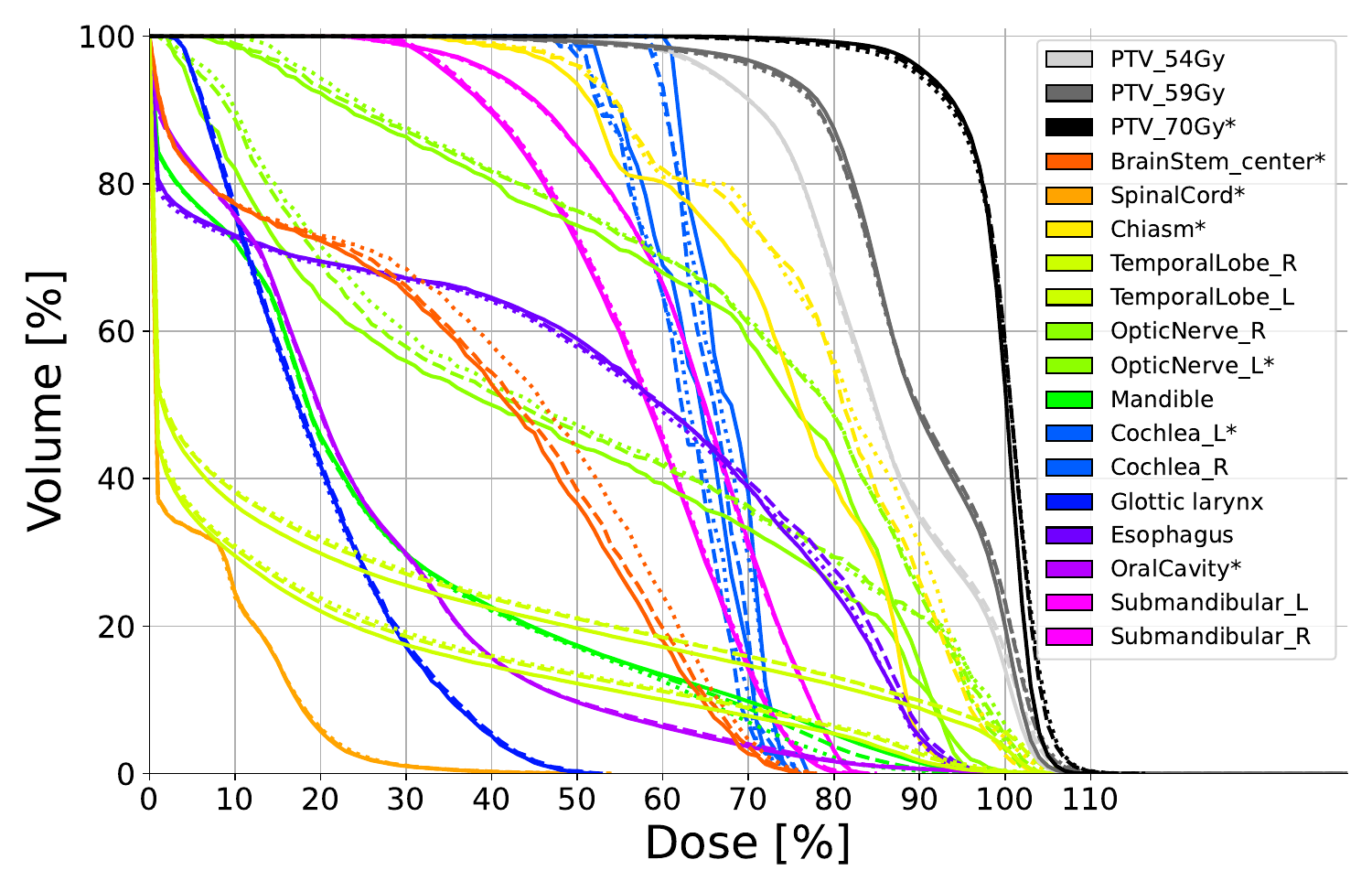}
    \label{fig:sub2}
  \end{subfigure}%
  \begin{subfigure}{0.5\textwidth}
    \includegraphics[width=\linewidth]{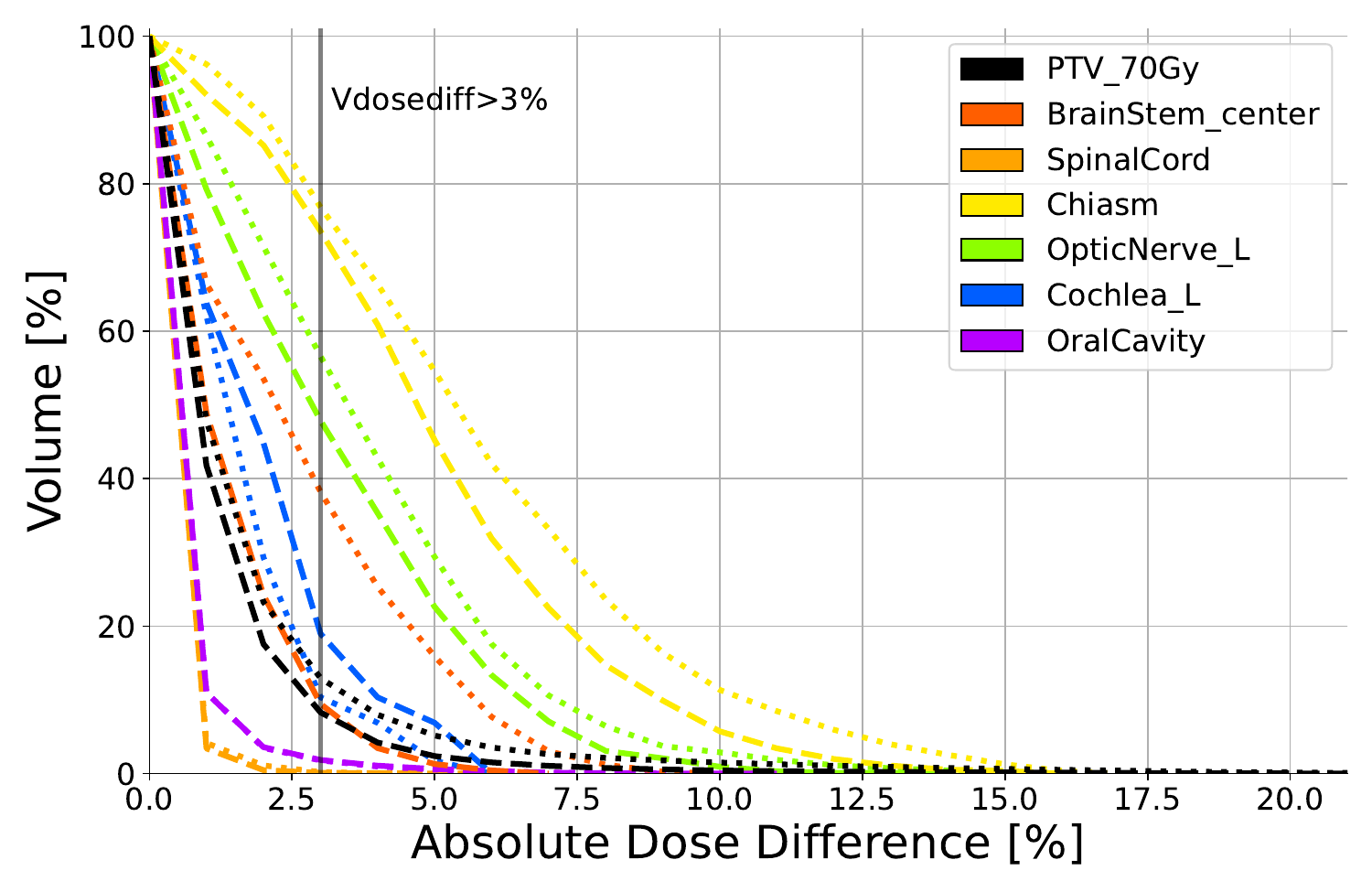}
    \label{fig:sub3}
  \end{subfigure}
   \vspace{-30pt}
  \caption{Dosimetric study of a single case. \textbf{(Top)} Comparison of dose distribution. dDose was calculated as sDose - pDose. \textbf{(Bottom Left)} DVHs, with structures highlighted in the DDVH graph denoted by (*). \textbf{(Bottom Right)} DDVHs. Representation for DVHs and DDVHs: solid lines (---) for pDose, dotted lines (...) for pix2pix model, dashdot (- - -) lines for DSBM (50-step).} 
  \label{fig:dose}
\end{figure*}

\begin{figure*}
  \begin{subfigure}{0.75\linewidth}
    \includegraphics[width=\linewidth]{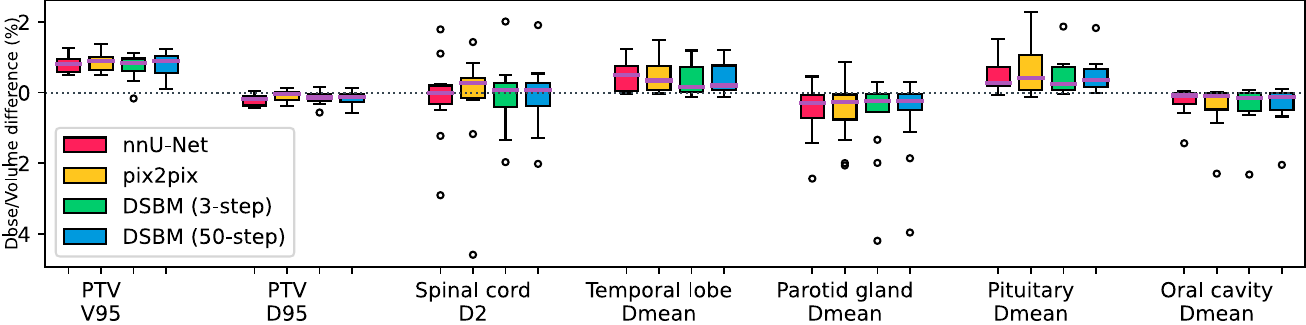}
  \end{subfigure}
    \begin{subfigure}{0.24\linewidth}
    \includegraphics[width=\linewidth]{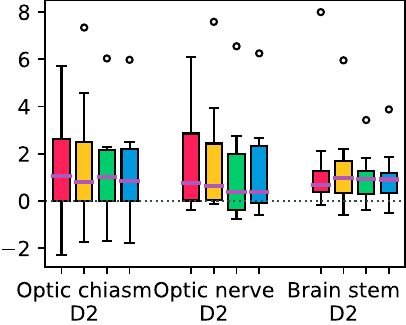}
    \end{subfigure}

  \begin{subfigure}{\linewidth}
    \centering
    \includegraphics[width=1.0\linewidth]{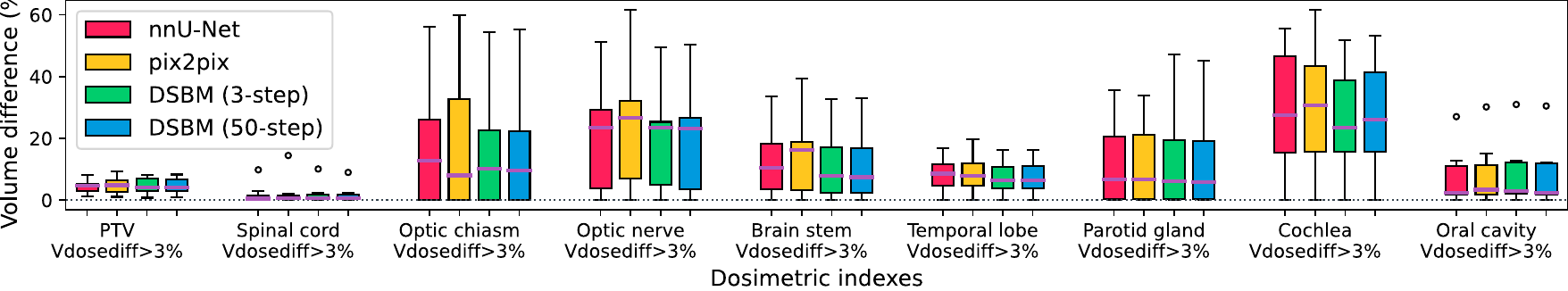}
  \end{subfigure}
  \vspace{-15pt}
  \caption{Comparison of the dosimetric accuracy of the sCTs (N=12), in terms of (\textbf{upper}) differences in typical dose indexes from corresponding DVHs, and (\textbf{bottom}) differences in Vdosediff>3\% from corresponding DDVH.}
  \label{fig:combined2}
\end{figure*}

At the dosimetric level, DSBM also demonstrated improved agreement for calculating doses when compared to those calculated on the corresponding pCT. It showed mean voxel-wise differences in the body region of less than 0.01\% of the prescribed dose (Table \ref{table:2}), and the 1\%/1mm gamma pass rate of 95.53 ± 2.89\% in the RoI (Table \ref{table:3}). Figure \ref{fig:dose} offers a visual comparison between synthetic (sDoses) and planned doses (pDose) in terms of voxel-wise dose differences, DVHs and DDVHs of a single case. Figure \ref{fig:combined2} compares the differences of dosimetric indices for PTVs and OARs across all 12 cases in the testing set. These results underscore DSBM's dosimetric accuracy, confirming its efficacy and potential utility in clinical proton therapy settings.

\begin{table}[t]
\begin{center}
\setlength{\tabcolsep}{1.8pt}
\caption{Quantitative comparisons between results from DSBM (using different number of neural function evaluation steps) and conventional approaches.}
\footnotesize
\vspace{-5pt}
\begin{tabular}{l|ccc|c}
\toprule
\textbf{Method} & \textbf{MAE (HU)} & \textbf{PSNR (dB)} & \textbf{SSIM (\%)} & \textbf{Dice\_bone (\%)} \\ \hline
nnU-Net                                   & 80.44±9.02                                  & 27.12±1.09                                   & 80.52±2.72                                   & 81.55±4.00                                         \\
pix2pix                                   & 77.41±7.81                                  & 27.50±0.89                                   & 81.95±2.68                                   & 80.72±3.74                                         \\ \hline
DSBM (1-step)                             & 73.00±8.41                                  & 28.16±1.07                                   & 83.09±2.72                                   & 82.95±3.92                                         \\
DSBM (3-step)                             & \textbf{72.30±8.31}                         & \textbf{28.20±1.07}                          & \textbf{83.26±2.69}                          & \textbf{83.10±3.90}                                \\
DSBM (10-step)                            & 72.64±8.25                                  & 28.19±1.05                                   & 83.21±2.71                                   & 82.99±3.90                                         \\
DSBM (50-step)                            & 73.29±8.28                                  & 28.13±1.05                                   & 83.13±2.72                                   & 82.83±3.89                                         \\ \bottomrule
\end{tabular}
\label{tab:1}
\vspace{-20pt}
\end{center}
\end{table}

\begin{table}[t]
\begin{center}
\setlength{\tabcolsep}{1.3pt}
\caption{Mean Dose Error (\%) from different methods by different thresholds.}
\small
\vspace{-5pt}
\begin{tabular}{l|cccc}
\toprule
\textbf{Method} & \textbf{Dose\_body} & \textbf{Dose\textgreater{}10\%} & \textbf{Dose\textgreater{}50\%} & \textbf{Dose\textgreater{}90\%} \\ \hline
nnU-Net         & 0.04±0.05           & 0.13±0.09                       & 0.39±0.22                       & 0.40±0.25                       \\
pix2pix         & 0.02±0.05           & 0.08±0.10                       & 0.32±0.23                       & 0.37±0.25                       \\ \hline
DSBM (1-step)   & -0.01±0.07          & \textbf{0.06±0.13}              & \textbf{0.26±0.32}              & \textbf{0.31±0.32}              \\
DSBM (3-step)   & \textbf{0.00±0.07}  & 0.08±0.12                       & 0.29±0.31                       & 0.33±0.30                       \\
DSBM (10-step)  & 0.01±0.06           & 0.09±0.11                       & 0.31±0.29                       & 0.35±0.29                       \\
DSBM (50-step)  & 0.01±0.06           & 0.09±0.11                       & 0.32±0.29                       & 0.36±0.28                       \\ \bottomrule
\end{tabular}
\label{table:2}
\vspace{-25pt}
\end{center}
\end{table}

\begin{table}[t]
\begin{center}
\setlength{\tabcolsep}{17pt}
\caption{Gamma pass rate (\%) of sDoses over different thresholds.}
\small
\vspace{-5pt}
\begin{tabular}{l|cc}
\toprule
\textbf{Method} & \textbf{1\%/1mm}    & \textbf{2\%/1mm}    \\ \hline
nnU-Net       & 95.28±3.19          & 96.01±2.96          \\
pix2pix       & 95.42±2.95          & 96.15±2.72          \\ \hline
DSBM (1-step)   & 95.47±2.99          & 96.18±2.77          \\
DSBM (3-step)   & 95.51±2.93          & 96.22±2.70          \\
DSBM (10-step)  & 95.53±2.91          & \textbf{96.23±2.68}          \\
DSBM (50-step)  & \textbf{95.53±2.89} & 96.22±2.70 \\ \bottomrule
\end{tabular}
\label{table:3}
\vspace{-20pt}
\end{center}
\end{table}
\section{Discussion}
The evaluations in our study, both at the image and dosimetric levels, have demonstrated the superiority of the Diffusion Schrödinger Bridge Model (DSBM) approach over conventional classic MR-CT synthesis methods. The DSBM-generated CT images not only exhibit remarkable accuracy and anatomical fidelity but also demonstrate dosimetric accuracy, vital for applications like adaptive or even MR-only proton treatment planning. A notable aspect of the DSBM method is its efficiency in generating high-quality results even with a limited number of neural function evaluation steps (NFEs). Our findings reveal that the DSBM can still outperform conventional methods when operating in very few sampling steps, even in a single step. In practical terms, generating a volume batch of size $128\times128\times4$ pixels takes only about 0.2 seconds per sampling step with DSBM. It should be noticed that conventional diffusion models require over 100 sampling steps for comparable results, which can lead to a generation time of more than 20 seconds per volume batch. Consequently, the DSBM method offers a speed enhancement of 10 to 50 times, marking a substantial leap in efficiency. Moreover, increasing the number of sampling steps does not necessarily correlate with further outcome improvement. We postulate that this phenomenon could be attributed to the introduction of noise at each step due to the inherent randomness in the sampling process.

\vspace{-5pt}
\section{Conclusion}
In this study, we propose the Diffusion Schrödinger Bridge Models (DSBM) which enhances the performance of MR-to-CT image synthesis, both in quality and efficiency. By learning the nonlinear diffusion processes between MR and CT data distributions and initiating synthesis from a prior distribution, DSBM outperforms traditional methods, as validated in our head and neck cancer case studies. This approach has demonstrated its value through comprehensive evaluations at both the image and dosimetric levels. DSBM can improve the accuracy and efficiency of MR-based treatment planning in proton therapy, which could potentially reduce imaging-related dose and improve patient outcomes in radiotherapy.













\printbibliography

\end{document}